\documentclass[letterpaper]{article}
\usepackage{iccc}
\usepackage{fancyhdr}
\usepackage{graphicx}

\RequirePackage{natbib}
\renewcommand\cite{\citep}
\renewcommand\shortcite{\citeyearpar}

\usepackage{times}
\usepackage{helvet}
\usepackage{courier}
\usepackage{caption}
\usepackage[table,xcdraw]{xcolor}  
\usepackage{array}  

\usepackage{tabularx}  
\usepackage{multirow}

\usepackage[colorlinks=true, linkcolor=blue, citecolor=blue, urlcolor=black]{hyperref}
\usepackage{xurl}
\title{Ways of Seeing, and Selling, AI Art}
\author{Imke van Heerden\\
Department of Digital Humanities\\
King's College London\\
Strand, London WC2R 2LS, UK\\
imke.vanheerden@kcl.ac.uk\\
}
\begin{document} 
\maketitle
\thispagestyle{plain}
\begin{abstract}
In early 2025, \textit{Augmented Intelligence} -- Christie’s first AI art auction -- drew criticism for showcasing a controversial genre. Amid wider legal uncertainty, artists voiced concerns over data mining practices, notably with respect to copyright. The backlash could be viewed as a microcosm of AI’s contested position in the creative economy. Touching on the auction’s presentation, reception, and results, this paper explores how, among social dissonance, machine learning finds its place in the artworld. Foregrounding responsible innovation, the paper provides a balanced perspective that champions creators’ rights and brings nuance to this polarised debate. With a focus on exhibition design, it centres \textit{framing}, which refers to the way a piece is presented to influence consumer perception. Context plays a central role in shaping our understanding of how good, valuable, and even ethical an artwork is. In this regard, \textit{Augmented Intelligence} situates AI art within a surprisingly traditional framework, leveraging hallmarks of ``high art'' to establish the genre’s cultural credibility. Generative AI has a clear economic dimension, converging questions of artistic merit with those of monetary worth. Scholarship on ways of seeing, or framing, could substantively inform the interpretation and evaluation of creative outputs, including assessments of their aesthetic and commercial value.
\end{abstract}

\vspace{5mm}

\begin{flushright}
\footnotesize{\textit{Peter Fuller: You can’t sell the Sistine Chapel.}} \\
\footnotesize{\textit{Questioner: They would if they could.}}\footnote{This excerpt was sourced from an edited transcript of an open discussion between critics Terry Eagleton and Peter Fuller at the Institute of Contemporary Arts \shortcite{eagleton1983question}.}
\end{flushright}

\section{Introduction}

\begin{figure}[t]
\includegraphics[width=\columnwidth]{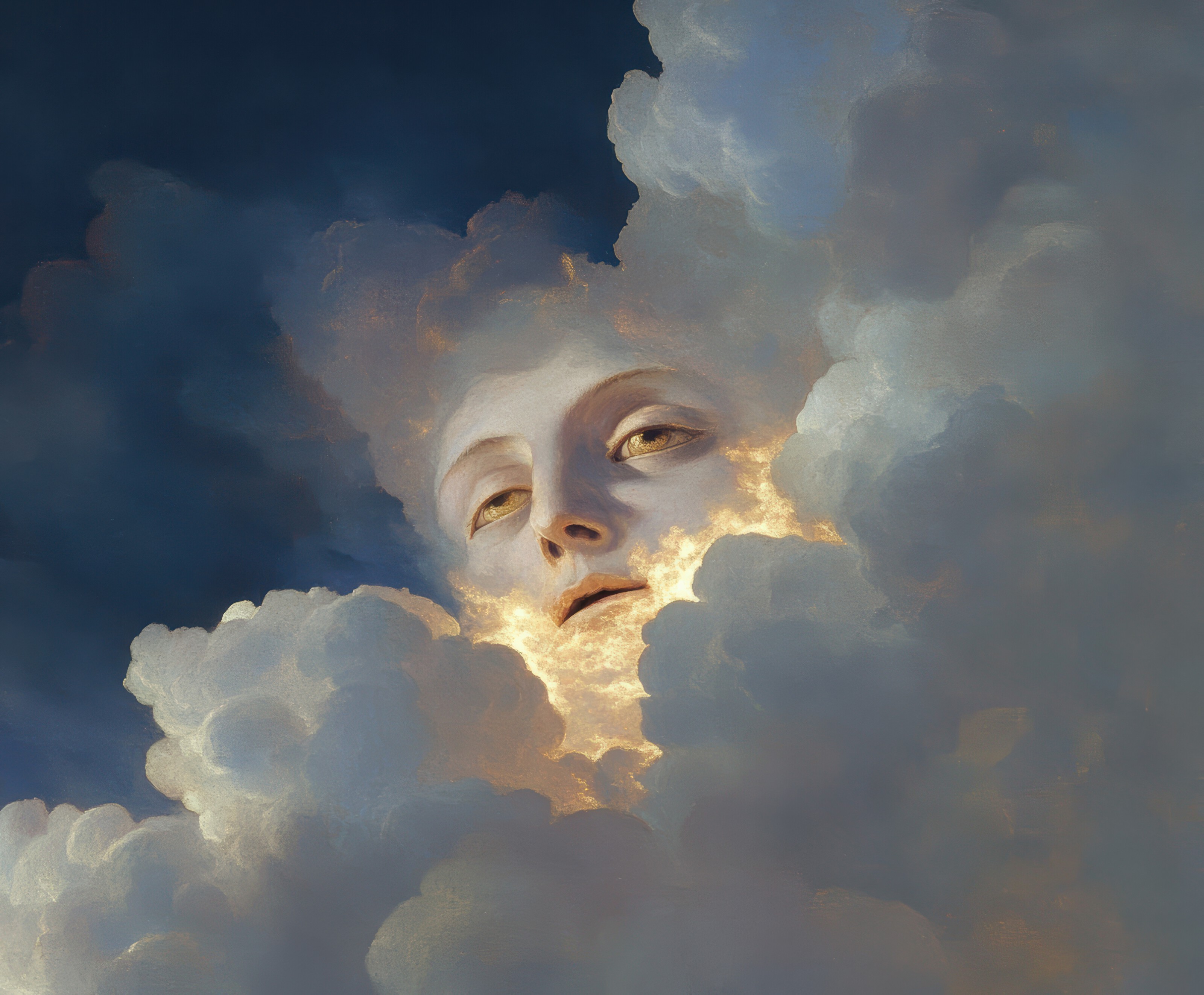}
\caption{``Golden Breath'' by Keke © Dark Sando / Fellowship}
\label{fig:keke}
\vspace{-4mm}
\end{figure}

In ``Golden Breath'', a divine being awakens slowly to its power. With eyes reminiscent of Michelangelo’s David or, perhaps, the Sistine Chapel \textit{ignudi}, this figure could have been painted by a Renaissance artist (as shown in \autoref{fig:keke}). The image, however, was generated by an autonomous entity described by its developer Dark Sando as ``an artist with dark, captivating sensibilities''.\footnote{The artwork's description is provided on Christie’s website at \href{https://onlineonly.christies.com/s/augmented-intelligence/keke-20/250104}{https://onlineonly.christies.com/s/augmented-intelligence/keke-20/250104}.}

What is particularly intriguing about the work is the pairing of a digital piece with a physical painting: the picture was first produced by the AI model, Keke, and then hand-painted by The Sable Collective, a group of UK-based painters led by Nicolas Archer. This human-machine collaboration resulted in an acrylic and oil-on-linen, which was among the works featured in Christie’s first AI art auction.\footnote{For additional information, please refer to the auction webpage available at \href{https://onlineonly.christies.com/s/augmented-intelligence/lots/3837}{https://onlineonly.christies.com/s/augmented-intelligence/lots/3837}.} 

\textit{Augmented Intelligence} ran from February 20\textsuperscript{\scriptsize th} to March 5\textsuperscript{\scriptsize th}, 2025 and included a diverse range of artists, from computer art pioneers Harold Cohen and Charles Csuri to contemporary artists like Sougwen Chung and Claire Silver. The auction, however, had launched amid significant controversy. Approximately 6,500 people, including visual artists, signed a well-publicised open letter\footnote{The letter is available at \href{https://openletter.earth/cancel-the-christies-ai-art-auction-f5135435}{https://openletter.earth/cancel-the-christies-ai-art-auction-f5135435}.} calling for its cancellation but, despite their efforts, bidding opened as planned. The letter expresses concern over the potential use of copyrighted material to train generative models, claiming that the auction encourages ``AI companies’ mass theft of human artists’ work''. Renowned media artist Refik Anadol, whose auctioned piece drew from 1.2 million images captured by the International Space Station and satellites, linked the reaction to ``lazy critic practices and doomsday hysteria'' \cite{Milmo2025}.

\section{AI, Art, and Ethics}

The backlash against \textit{Augmented Intelligence} has spoken to the wider conversation on artists’ rights. Serious points of contention include copyright, consent, compensation, and attribution, with groups of visual artists taking legal action against Stability AI \cite{Brittain2023} and Google \cite{Soni2024}. In the UK, tensions are running high following the Government’s open consultation on copyright and artificial intelligence. Concurrent with (though not directly related to) the auction, prominent creatives \shortcite{TheTimes2025}, including Sir Stephen Fry, Sir Paul McCartney, and Dua Lipa, have written to \textit{The Times} to oppose the UK government’s proposal for a new copyright exemption. The joint letter states that ``Britain’s creative industries want to play their part in the AI revolution, as they have with new technologies in the past. But if this is to succeed, they need to do so from a firm intellectual-property base''.

While AI-assisted art is often viewed through the lens of publicly available models, which rely on massive quantities of existing media, there are artists who develop in-house tools, gather their own datasets, or use licensed resources to train AI models. As one example from the auction, Sofia Crespo and Anna Ridler created ``Long Short-Term Memories (sketches from a garden in Argentina)'' by using photographs they personally captured on trips to South America.

In the public sphere, the auction has functioned productively to connect theoretical discussions on AI and creativity with concrete examples. When different art-making processes are painted with the same brush, we cannot gain insight into this critical issue. Dismissing all computer-generated works solely on principle overlooks creatives’ own attempts to exercise agency and shape AI in fair and ethical ways. There is a diversity of approaches in computational creativity – and some are more responsible than others. On this point, it should be emphasised that critical and imaginative reflections do not diminish the validity of artists’ concerns; in periods of heightened urgency, nuanced responses may support long-term solutions.

Attentive to the gravity and complexity of the situation, the aim of this position paper is not to pass judgment on how the auction’s respective pieces were created. Beyond individual artworks, tensions have animated broader questions of aesthetic and economic value with longstanding histories. Bigger questions include: How does such a controversial genre demonstrate belonging in a precarious climate? How does it cultivate insider status? By which strategies are we, as viewers and potential buyers, persuaded to find authenticity in the artificial? Rob Horning’s reflection \shortcite{Horning2017} is particularly compelling in this respect, that a ``rhetoric of authenticity comes to the fore when what could be true seems especially vulnerable, debatable, difficult, riven, up for auction or appropriation''. 

\begin{figure}[ht]
\includegraphics[width=\columnwidth, trim=0 45 0 45, clip]{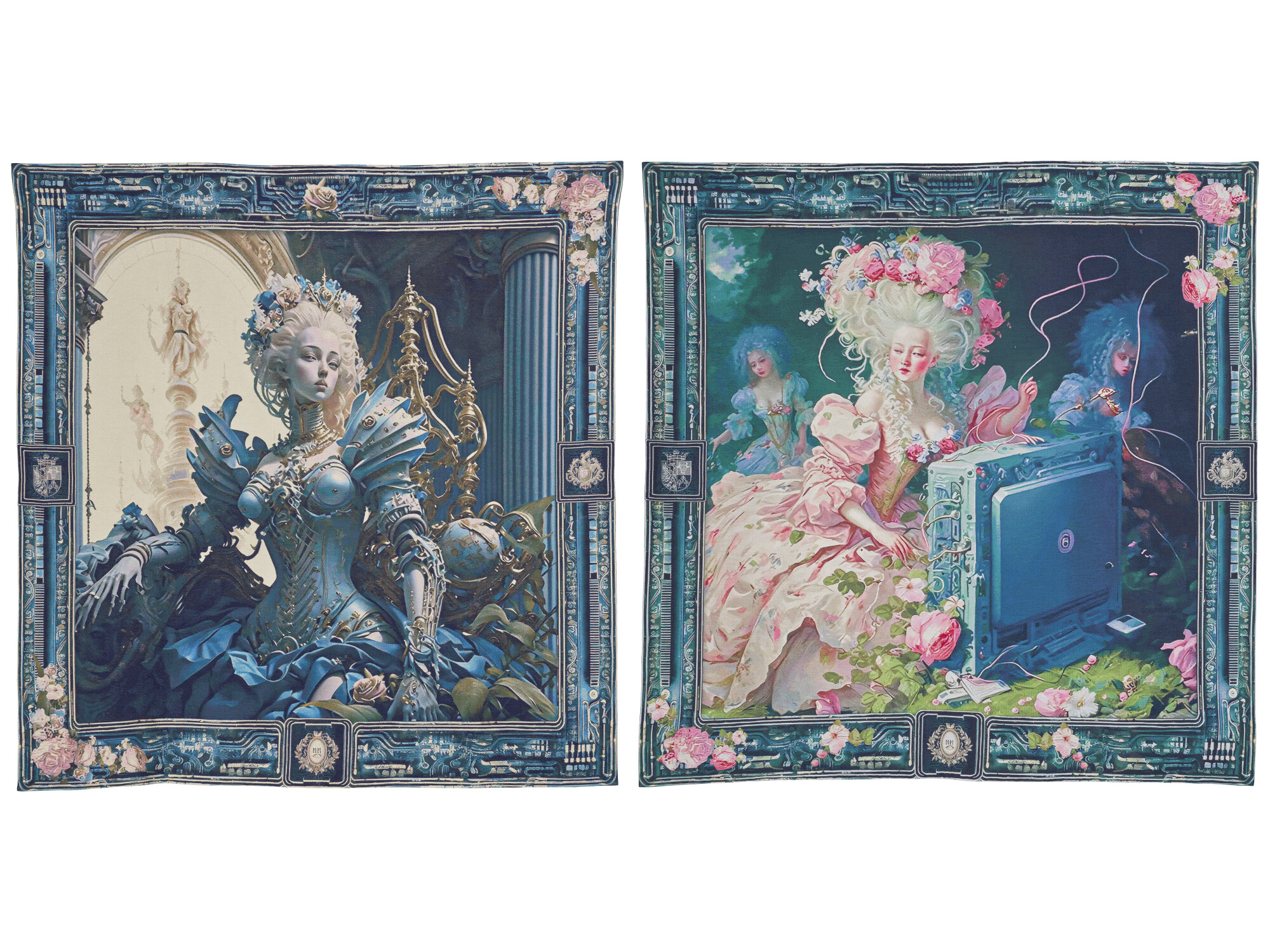}
\caption{``Marie Antoinette After the Singularity \#1'' and ``Marie Antoinette After the Singularity \#2'' © Grimes, Mac Boucher, Mariya Jacobo, Eurypheus / Christie’s Images Ltd.}
\label{fig:marie}
\end{figure}

\section{Selling AI Art}

Despite creators’ concerns, \textit{Augmented Intelligence} totalled \$728,784 in auction sales. \autoref{tab:list} provides a comprehensive list of the artworks sold, alongside corresponding prices. Refik Anadol received the highest price, of \$277,200, for his AI-driven data painting ``Machine Hallucinations -- ISS Dreams -- A''. Anadol's work has showcased in numerous high-profile events like the 2025 World Economic Forum in Davos and the AI Action Summit in Paris. Other artists included in the auction have featured in prestigious venues as well. In London, for instance, Jake Elwes has exhibited in Somerset House; Holly Herndon and Mat Dryhurst, in the Serpentine Gallery; and Harold Cohen, in Tate Modern.

AI, thus, has achieved a certain status in the art world. This raises the question of why some pieces are deemed acceptable, or actually admired, against a critical backdrop. When does artistic licence warrant forgiveness and when does it provoke ire? Where do we draw the line between fake and fantastic? And what degree of human involvement is required to negotiate such thresholds? These questions broach the aforementioned ethical issues (such as copyright infringement) but also involve aesthetic and philosophical considerations. Furthermore, creative AI has an undeniable economic overlay. Research in this domain could better discern the financial challenges, motivations, and implications surrounding creative technologies.

To this end, we can examine how, and under which conditions, AI art gains recognition and legitimacy. In marketing theory, the concept of \textit{framing} \cite{goffman1974frame} refers to how information -- or, in this case, a work of art -- is positioned within specific sociocultural and institutional contexts. Extending to exhibition design, curatorial decisions, and promotional content, framing strategies highlight certain attributes in order to resonate with consumers. That is to say, extrinsic factors have a bearing on estimations of a work's intrinsic value. According to Danto \shortcite{Danto1964}, ``telling artworks from other things is not so simple a matter [...] and these days one might not be aware [one] was on artistic terrain without an artistic theory to tell [one] so''. In this regard, it appears that \textit{Augmented Intelligence} gives prominence to physicality, provenance, and performance.

\onecolumn
\begin{table*}[ht]
\centering
\renewcommand{\tabularxcolumn}[1]{m{#1}}
\begin{tabularx}{\textwidth}{|m{4.7cm}|X|c|c|}
\hline
\multicolumn{1}{|c|}{\multirow{2}{*}{\textbf{Artist(s)}}} & 
\multicolumn{1}{c|}{\multirow{2}{*}{\textbf{Lot}}} &
\multicolumn{2}{c|}{\textbf{Price (USD)}} \\ 
\cline{3-4}
\multicolumn{1}{|c|}{} & \multicolumn{1}{c|}{} & \textbf{Estimated} & \textbf{Realised} \\ 
\hline
Refik Anadol & \textit{Machine Hallucinations -- ISS Dreams -- A} & 150,000 – 200,000 & 277,200 \\ \hline
Holly Herndon and Mat Dryhurst & \textit{Embedding Study 1 \& 2 \newline (from the xhairymutantx series)} & 70,000 – 90,000 & 94,500 \\ \hline
Charles Csuri & \textit{Bspline Men} & 55,000 – 65,000 & 50,400 \\ \hline
Claire Silver & \textit{daughter} & 40,000 – 60,000 & 44,100 \\
\hline
Jesse Woolston & \textit{The Dissolution Waiapu} & 30,000 – 50,000 & 40,320 \\ \hline
Sougwen Chung & \textit{Study 33} & 20,000 – 30,000 & 25,200 \\ \hline
Grimes, Mac Boucher, \newline Mariya Jacobo, Eurypheus & \textit{Marie Antoinette After the Singularity \#1 \& \#2} & 40,000 – 60,000 & 25,200 \\ \hline
Keke & \textit{Golden Breath} & 15,000 – 20,000 & 21,420\\ \hline
Scott Eaton & \textit{Human Allocation of Space} & 8,000 – 12,000 & 16,380\\ \hline
Owen Mcateer & \textit{Seiche} & 12,000 – 18,000 & 12,600\\ \hline
Gene Kogan & \textit{Latent Character Arithmetic: \newline King - Male + Female = ``Queen''} & 6,000 – 8,000 & 12,600 \\ \hline
Harold Cohen & \textit{Untitled (i23-3758)} & 10,000 – 15,000 & 11,340 \\ \hline
Niceaunties & \textit{5 Mins To Opening} & 12,000 – 18,000 & 11,340 \\ \hline
Brendan Dawes & \textit{Altarpiece: The Divinity} & 8,000 – 10,000 & 10,710 \\ \hline
Sofia Crespo and Anna Ridler & \textit{Long Short Term Memories \newline (sketches from a garden in Argentina)} & 15,000 – 20,000 & 10,080\\ \hline
Sasha Stiles & \textit{\small{WORDS CAN COMMUNICATE BEYOND WORDS}} & 10,000 – 15,000 & 10,080\\ \hline
Linda Dounia & $\textit{14}^\circ\: \textit{40}'\: \textit{34.46}''$ \textit{N} $\textit{17}^\circ\: \textit{26}'\: \textit{15.14}''$ \textit{W} & 4,000 – 6,000 & 9,450 \\ \hline
Daniel Ambrosi & \textit{Central Park Nightfall} & 7,000 – 10,000 & 9,450\\ \hline
Sarp Kerem Yavuz & \textit{Hayal (``Dream'')} & 8,000 – 12,000 & 8,820\\ \hline
Alexander Reben & \textit{Untitled Robot Painting, 2025} & 100 – 1,728,000 & 8,190 \\ \hline
Alkan Avcioğlu & \textit{Liquid Modernity} & 2,000 – 4,000 & 5,040\\ \hline
Ganbrood & \textit{Gynaeceum} & 5,000 – 10,000 & 3,276\\ \hline
Kevin Abosch & \textit{NOVEL ENCRYPTION \#5} & 3,000 – 5,000 & 2,772\\ \hline
Roope Rainisto & \textit{The Swimming Hall} & 2,000 – 4,000 & 2,016\\ \hline
Deepblack & \textit{DeepBlack \#100} & 1,000 – 2,000 & 2,016 \\ \hline
Clownvamp & \textit{JUNK \#8} & 1,000 – 2,000 & 1,764\\ \hline
Vanessa Rosa & \textit{Little Martians \& Abraham} & 2,000 – 4,000 & 1,764\\ \hline
Amelia Winger-Bearskin & \textit{Nearest Neighbor} & 5,000 – 7,000 & 756\\ \hline
Botto & \textit{Siamese Cycle in Absurdism} & 20,000 – 30,000 & n/a\\ \hline
Huemin & \textit{Dream-0 \#9} & 30,000 – 50,000 & n/a\\ \hline
Ivona Tau & \textit{Nightcall (Not AI)} & 7,000 – 10,000 & n/a\\ \hline
Jake Elwes & \textit{Zizi -- Queering the Dataset (07)} & 18,000 – 25,000 & n/a\\ \hline
Pindar Van Arman & \textit{Emerging Faces} & 180,000 – 250,000 & n/a\\ \hline
Robbie Barrat and Ronan Barrot & \textit{Infinite Skull \#21} & 10,000 – 15,000 & n/a\\ \hline
\end{tabularx}
\caption{The full list of auctioned artworks are arranged in descending order of the final price achieved. Prices are provided in US Dollar, but cryptocurrency payments were also accepted. Corresponding pre-sale estimates are included as well. ``Not applicable'' indicates instances where a bid was not reported. Data was sourced in March 2025 from Christie’s official website: \href{https://onlineonly.christies.com/s/augmented-intelligence/lots/3837}{https://onlineonly.christies.com/s/augmented-intelligence/lots/3837}.}
\label{tab:list}
\end{table*}
\twocolumn

\begin{figure}[ht]
\includegraphics[width=\columnwidth]{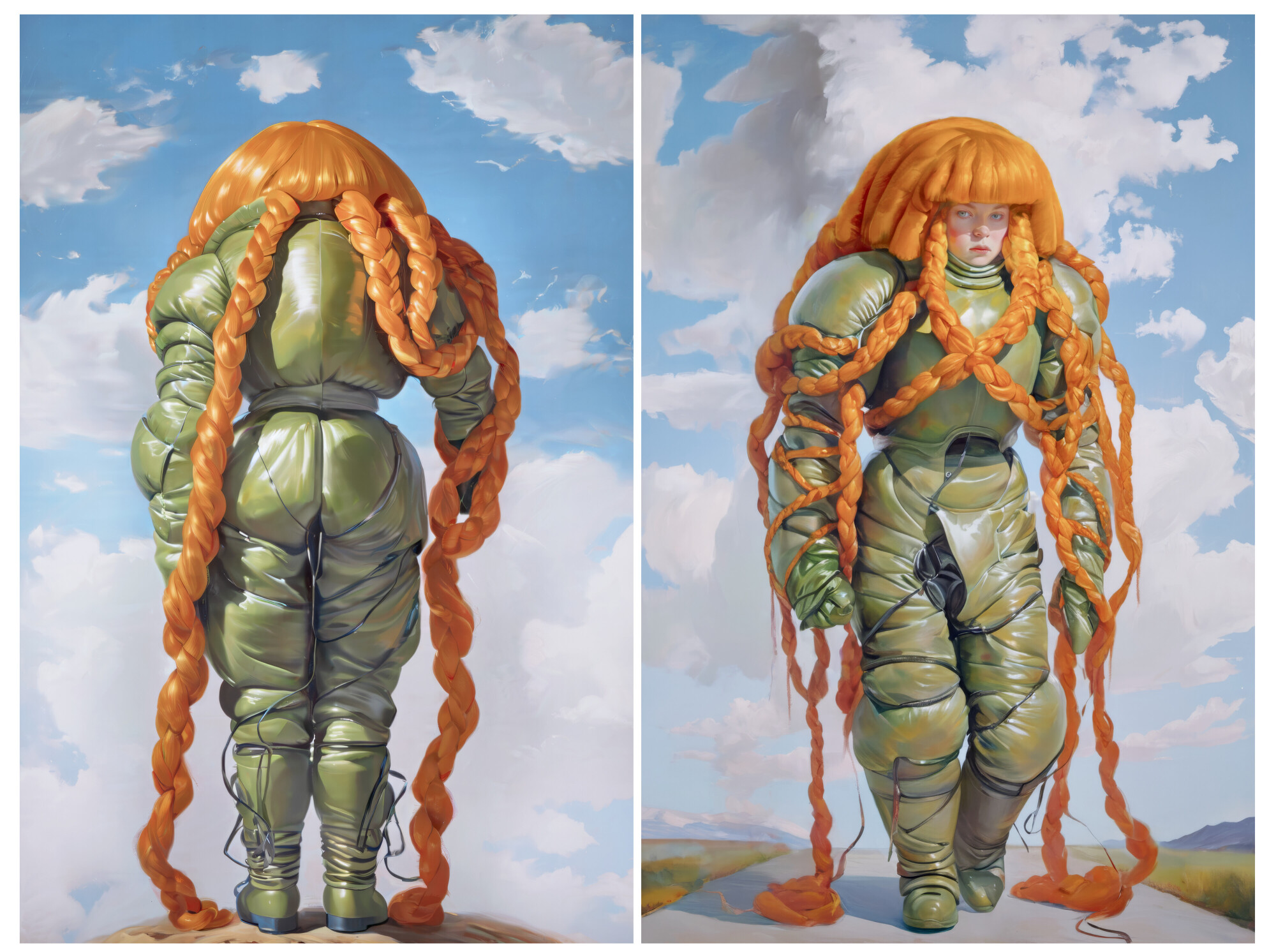}
\caption{``Embedding Study 1 \& 2'' © Holly Herndon and Mat Dryhurst / Christie’s Images Ltd.}
\label{fig:embedding}
\end{figure}
\vspace{-6mm}

\subsection{The Material (Re)Turn}
Although the majority of \textit{Augmented Intelligence}’s lots consist of non-fungible tokens (NFTs), many include physical works as well. It is in this regard that we discover a surprising array of analogue media and techniques, such as oil painting, ink on paper, weaving, handmade ceramic, sculpture, and Polaroid photography. The auction was accompanied by an exhibition at Christie’s New York headquarters. Artworks were mounted in a gallery space, including ``Embedding Study'', twin prints by Holly Herndon and Mat Dryhurst exhibited in vibrant colour at near-life size (as shown in \autoref{fig:embedding}). Another pair, ``Marie Antoinette After the Singularity'' (\autoref{fig:marie}), vividly demonstrates the interplay of old and new, highlighting the work’s synthetic origins. These Rococo-style, jacquard-woven tapestries are based on generated images of a cyborg Marie Antoinette.\footnote{In technological history, the jacquard loom is often seen as an early symbol of human-computer interaction.} 

We must keep in mind that most of the auction’s pieces are born digital, which means they were initially created and exist primarily on screen. It is an obvious advantage to have a physical replica, for display in a real-world environment, but there is another reason why digital media would be accompanied by tangible counterparts. It can be challenging to establish provenance in a domain of endless sharing and editing, where images are often estranged from their creators and contexts. Generative platforms, in particular, can be prompted to produce infinite variations on a theme. 

Thinking back to Walter Benjamin’s influential essay, ``The Work of Art in the Age of Mechanical Reproduction'' \shortcite{benjamin}, I’d say that blockchain verification cannot convey the same sense of heritage as a physical artifact: an object you can hold in your hands, with a history you can feel. Underpinning the auction is the view that art is purchased for its rarity; in material form, an artwork may lay claim to the substance and authenticity of an original, a permanent record that cannot be altered.

\subsection{Authorship}

In addition to physical presence, which is a traditional marker of value, \textit{Augmented Intelligence} foregrounds authorship. Readers have grown accustomed to headlines like ``AI creates art'', which credits the technology. In contrast, the auction prioritises human agency. Typically, each lot is prominently headed by the artist’s name and year of birth – as opposed to, say, an AI model or platform. This appears to be the standard layout of Christie’s website, but the structure is nonetheless revealing, illustrating how AI slots into existing aesthetic frameworks like authorial intent.

To clarify, the choice may seem straightforward when, in reality, generative art vastly complicates artistic authorship and ownership \cite{van2021ai}. Should authorship be ascribed to the human artist, the AI model, or both? What about the computer scientists and software developers behind the algorithms? Moreover, how can we acknowledge the creators of the work the model was trained on? Don’t they deserve recognition as well? \textit{Augmented Intelligence} places the human artist front and centre, largely presenting AI as neither superintellingent nor sentient, but as a tool that enhances human creativity. Mindful of ethics, it can be beneficial and, in some cases, certainly accurate to credit the human artist’s vision, which encourages accountability as well. However, there is an additional financial benefit in removing ambiguity around authorship. The presence of a human artist is decisive not only in authenticating a digital work but also in framing it as \textit{art} and \textit{asset}.

\vspace{2mm}

\subsection{Performance}

\begin{figure}[t]
\includegraphics[width=\columnwidth]{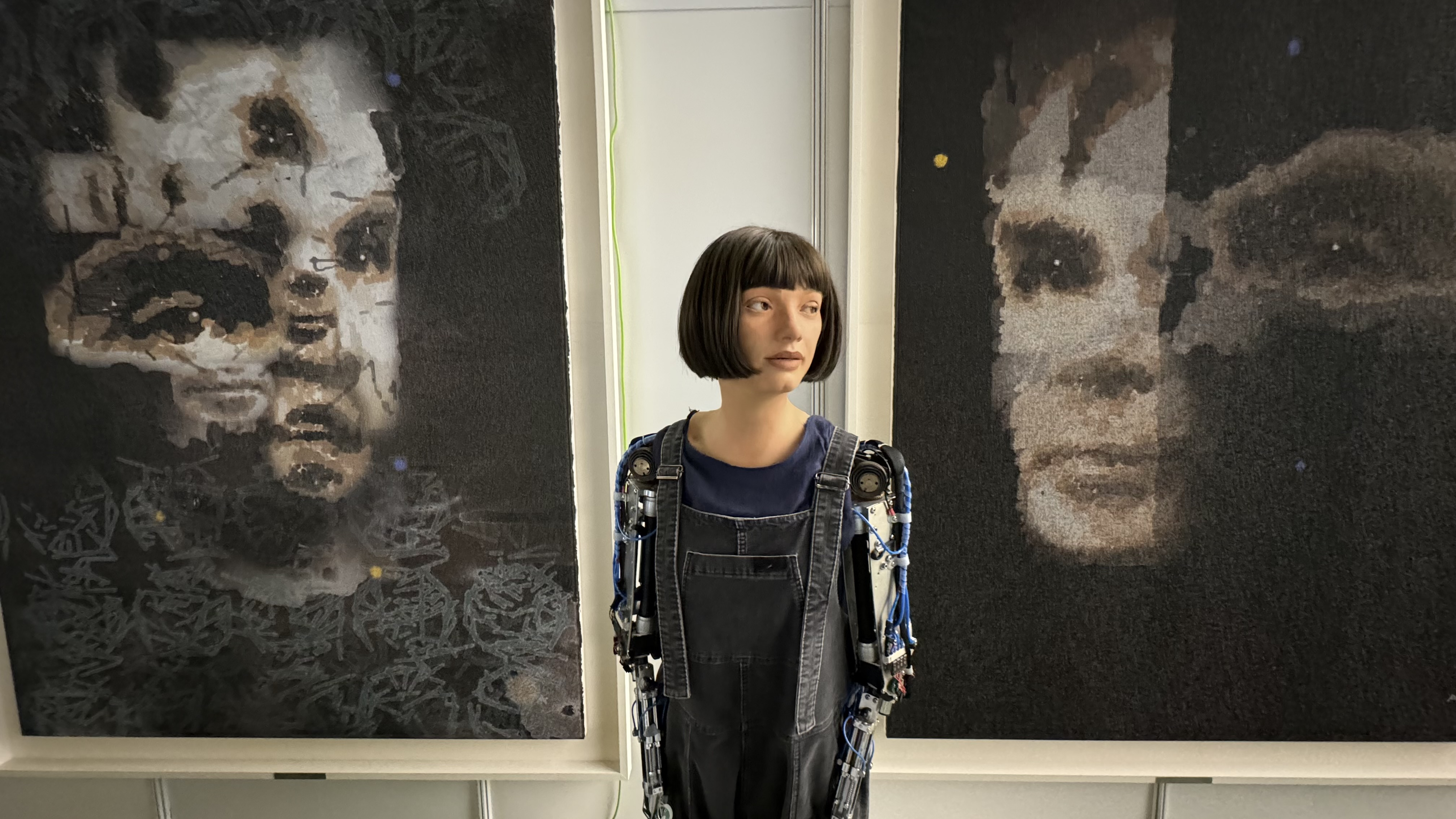}
\caption{Ai-Da at the United Nations, with ``A.I. God'' pictured right. © Ai-Da Robot Studios, \href{https://www.ai-darobot.com/}{www.ai-darobot.com}}
\label{fig:AiDa}
\end{figure}

\textit{Augmented Intelligence} followed on the heels of a landmark sale at Sotheby’s, where this trend was even more apparent. In November 2024, a humanoid robot’s portrait of Alan Turing sold for approximately \$1 million \cite{Roberts2024}. Titled ``A.I. God'', the portrait was unveiled at the United Nations' AI for Good Global Summit in Geneva last year as part of a large-scale, five-panelled Polyptych. Two of the panels are shown in \autoref{fig:AiDa}.

Ai-Da, the robot artist, uses eye cameras, a robotic arm, and AI algorithms to create physical paintings and drawings. Although ``A.I. God'' could be considered groundbreaking, given the manner of its creation, it does not break with convention. In fact, I would argue that the work’s success partially derives from its familiar frame, a set of anthropomorphising conventions that emphasise the robot’s physicality and artistic intent.

Ai-Da embodies artificial intelligence. Her (woman-presenting) humanoid form is a persona to which we can effortlessly assign authorship. As Leah Henrickson and Simone Natale \shortcite{Henrickson2022} remind us, presentation is everything: social cues like Ai-Da’s realistic appearance and clothing style suggest autonomy and creativity. Interviews with Ai-Da have imparted a distinct point of view, albeit problematically, lending an impression of meaningful engagement. Moreover, we can watch her draw, in a live demonstration of skill and artistry that her website associates with ``art historical developments in the twentieth century''.

Similarly, \textit{Augmented Intelligence}’s centrepiece was Alexander Reben’s ``Untitled Robot Painting, 2025''. For the duration of the two-week auction, the artwork was generated and painted in real time by a Matr Labs robot at Rockefeller Center. The piece was initially estimated at \$100 – \$1,728,000 since the robot only painted in response to actual bids. When a bid was placed, the artwork grew by one square inch per \$100. At the sale’s conclusion, it had raised a lacklustre \$8,190 and thus remained incomplete: a small imprint of a human hand on an otherwise empty canvas. As in Ai-Da’s case, technical sophistication blended with performance to bring a virtual work to life before our very eyes.

\section{Ways of Seeing}

Perhaps it is surprising that, in a futuristic context, both the Ai-Da team and \textit{Augmented Intelligence} engaged traditional standards of desirability. However, as John Berger’s seminal \textit{Ways of Seeing} \shortcite{berger2008ways} suggests, tradition can be leveraged to serve commercial interests. As Berger aptly observes: ``Publicity is, in essence, nostalgic. It must sell the past to the future. It cannot itself supply the standards of its own claims. And so all its references to quality are bound to be retrospective and traditional''. This remark provides an elegant explanation of why AI art may appeal to the past to create commercial value in the present.

To be sold as art, a piece must first convince the market that it is art. By tapping into cultural assumptions about what art is, how it is created, and what it should look like, auction houses and galleries can subtly shape ways of seeing – and selling. In the framing of \textit{Augmented Intelligence}, I find that traditional notions of authorship and art’s object status serve a legitimising function, creating the illusion of solid ground in a slippery context.

In this sense, several of the auction's artists represent their work in distinctly hybrid terms, problematising digital-physical and human-machine dichotomies in experimental ways. Sougwen Chung, one of TIME100’s Most Influential People in AI \shortcite{Chung2024}, explores new, hybrid systems of relation by painting in tandem with robots. Jesse Woolston's hybrid sculpture of steel and digital canvas ``embodies the tension between fluidity and rigidity, nature and technology, art and machine''. Modelled after the Waiapu River in New Zealand, ``The Dissolution Waiapu'' embraces ``the convergence of human craftsmanship and computational precision''. Comparably, Vanessa Rosa describes ``Little Martians \& Abraham'' as ``bridging the gap between traditional craftsmanship and AI-generated art''. Sasha Stiles portrays her piece, a sculpture of steel and light, as an engagement with ``the paradox of flesh and code that defines contemporary existence''.\footnote{Quotations in this paragraph are sourced from artists' descriptions of their lots at \href{https://onlineonly.christies.com/s/augmented-intelligence/lots/3837}{https://onlineonly.christies.com/s/augmented-intelligence/lots/3837}.}

As these descriptions demonstrate, framing is a vital component of an artwork's identity, meaning, and ability to inspire trust. In instances of co-creativity, where roles may lack definition, framing acquires heightened relevance. Artists stand at the forefront of framing responsible futures for AI. In this sense, they have a leading role to play in transforming, what James O'Sullivan \shortcite{o2019towards} calls, ``a dominant language of revolution'' into ``a language of evolution''.






\bibliographystyle{iccc}
\bibliography{ref.bib}

\end{document}